\documentclass[aps,amsmath,amssymb,twocolumn,floatfix,prb]{revtex4}
    \usepackage{graphicx}   \usepackage{bm}
    \usepackage{epsf,times} \usepackage{epsfig}
    \usepackage{dcolumn}    \usepackage{bm,bbm}
    \usepackage[dvips]{color}   \usepackage{subfigure}
\newcommand{\bra}[1]{\mbox{$\left\langle #1 \right|$}}
\newcommand{\ket}[1]{\mbox{$\left|#1\right\rangle$}}

\newcommand{\ignore}[1]{}
\newcommand{\bea}{\begin{eqnarray}} \newcommand{\eea}{\end{eqnarray}}
\newcommand{\beq}{\begin{equation}} \newcommand{\eeq}{\end{equation}}

\def\A{\mathcal{A}}                 \def\B{\mathcal{B}}
\def\openone{\leavevmode\hbox{\small1\kern-4.2pt\normalsize1}}
\begin{document}
\title{Interplay of topological order and spin glassiness in the toric code under random magnetic fields}

\author{Dimitris I. Tsomokos$^{1}$, Tobias J. Osborne$^{1,2}$, Claudio Castelnovo$^{3}$}
\affiliation{$^1$Department of Mathematics, Royal Holloway, University of London, Egham, TW20 0EX, UK \\
$^2$Wissenschaftskolleg zu Berlin, Wallotstra{\ss}e 19, 14193 Berlin, Germany \\
$^3$Rudolf Peierls Centre for Theoretical Physics and Worcester College, University of Oxford, 1 Keble Road, Oxford OX1 3NP, UK}
\date{\today}

\begin{abstract}
We analyze the toric code model in the presence of quenched disorder, which is introduced via different types of random magnetic fields. In general, close to a quantum phase transition between a spin polarized phase and a topologically ordered one, we find that increasing the amount of disorder favors the topological phase. For some realizations of disorder, topological order can be robust against arbitrarily strong magnetic fields. In the case of the toric code in a random $\pm h$ field, we show that the system exhibits a quantum phase transition to a spin glass phase in an appropriate dual variables description. The survival of topological order in the spin glass phase is directly related to the percolation properties of the rigid lattice in the Edwards-Anderson bimodal spin glass model. According to recent numerical results for this model [Phys. Rev. B \textbf{82}, 214401 (2010)], it is likely that the rigid lattice does not percolate and, as a result, a new intermediate quantum phase appears in the random-field toric code. In this intermediate quantum phase, topological order coexists with spin glassiness.
\end{abstract}

\maketitle


\section{Introduction}
Quantum systems can be dramatically influenced by disorder. In phenomena where Anderson localization~\cite{Anderson_1958} takes place, a weak random potential can completely localize a band of states that is otherwise conducting. For two-dimensional systems, in particular, accepted wisdom suggests that quantum phases are destroyed if disorder is strong enough~\cite{IQHE}.

In the present work we revisit this basic intuition in a spin lattice system exhibiting \emph{topological order} \cite{Wen_book}. Our main goal is to investigate how a nonlocal topologically ordered phase responds to local quenched disorder. To this end, we focus on an exactly solvable model that supports topological order, namely, the toric code \cite{Kitaev,Dennis,TQC_review}.

We first examine a particular case of variational tensor product approach to the ground state of the system~\cite{Chen2010}, dubbed \emph{wavefunction deformation} \cite{book_chapter}. In this case the ground state is analytically tractable and topological properties can be mapped out via the topological entropy \cite{Hamma2005,S_topo}. All ground state equal-time correlators are equivalent to those of a classical $2+0$ dimensional Ising model and, in the uniform-field case, there is a continuous transition between a $\mathbb{Z}_2$ topological phase and a spin polarized phase through a critical point that falls in the two-dimensional (2D) Ising universality class~\cite{Castelnovo_Chamon,Stefanos}.

In the second part we examine a \emph{Hamiltonian deformation}, which is closer to physical intuition in that a random magnetic field is added to the toric code. This situation is analyzed via the conventional correspondence to a $2+1$ dimensional classical model, whose uniform-field critical point falls in the 3D Ising universality class.

We find that the addition of a random magnetic field to a uniform one favors in general the highly entangled phase with topological order over the spin polarized phase (with no long-range entanglement). As a result, we find that if a uniform magnetic field has driven the system into a spin polarized phase, switching on some disorder leads to a \emph{recovery} of topological order. This is an instance where the addition of local disorder increases the genuine many-body entanglement in the system.

In the case of a wavefunction deformation without a uniform field component, topological order survives for \emph{any amount} of disorder. This is not so for a Hamiltonian deformation, where an infinitely strong random field induces a non-topological polarized phase with spins pointing along the local field directions. However we argue that, in this case, an intermediate phase emerges and we establish an intriguing connection between the topological properties of this phase and the percolation properties of the so-called \emph{rigid lattice}~\cite{Barahona1982} in the Edwards-Anderson bimodal (EAB) spin glass model~\cite{EAB}. In particular, we show that if the rigid lattice of the 2D EAB spin glass model does not percolate, as conjectured recently in a preliminary numerical investigation \cite{Roma2010}, then, in the intermediate quantum phase, spin glass properties coexist with topological order. Due to the interplay and possible merging of topological and spin glass properties in this intermediate phase, we shall refer to it as a \emph{topological spin glass}.

In what follows we briefly review the toric code model and its basic properties (Sec. \ref{sec:II}) and then proceed to the main parts of the paper, in which we analyze the influence of random perturbations on the topological order of the toric code. In Sec. \ref{sec:III} we examine a wavefunction deformation, which leads to a fine-tuned model with a tractable ground state subspace. In this case, we are able to calculate the block entanglement and hence the topological entropy of the randomly perturbed toric code. In Sec. \ref{sec:IV} we examine a Hamiltonian deformation and establish the connection between percolation properties of the rigid lattice and topological properties of the random-field toric code model, in which a topological spin glass phase appears if the rigid lattice does not form a giant cluster spanning the system. In Sec. \ref{sec:V} we compare and contrast the two types of random perturbations to the toric code and develop a qualitative understanding of the stability of topological order against them. In the final section our results are summarized and some derivative open questions are discussed.


\section{Preliminaries: Toric Code Model and Topological Order \label{sec:II}}
Quantum states possessing topological order, such as those encountered in the quantum Hall effect, do not break any local symmetry \cite{Wen_book}. An example of such a system is the toric code \cite{Kitaev}, which is defined on a square lattice with periodic boundary conditions in both directions and spins-$1/2$ on its bonds. The Hamiltonian of the model,
\bea
H_{\rm TC} = - \lambda_{\rm A} \sum_{s} A_{s} - \lambda_{\rm B} \sum_{p} B_{p},
\label{H_TC}
\eea
is composed of a sum over the \emph{star} operators
\bea
A_{s} = \prod_{l \in {s}} \hat{\sigma}^{\rm x}_{l},
\eea
which are applied on bonds meeting on a vertex $s$; and the \emph{plaquette} operators
\bea
B_{p} = \prod_{l \in {p}} \hat{\sigma}^{\rm z}_{l},
\eea
applied on bonds forming a square $p$.

There is a fourfold degeneracy (on the torus) in the ground state corresponding to different topological sectors. If we denote by $w_1$ and $w_2$ the incontractible Wilson loops along the vertical and horizontal directions, respectively, the ground state manifold is spanned by
\bea
\ket{\Psi_{ij}} \equiv \{w_1^i w_2^j\ket{\Psi_{\rm TC}}; \; i,j=0,1\}
\eea
where
\bea
\ket{\Psi_{\rm TC}} = |G|^{-1/2} \sum_{g \in G} g \ket{\bf{0}}. \label{Psi_0_TC_clean}
\eea
Here $g \in G$ is a string of star operators, and $\ket{\bf{0}}$ is a reference state in which all spins are pointing in the $+Z$ direction. The star operators form an Abelian group $G$ of order $|G| = 2^{L^2-1}$ for an $L \times L$ lattice $\cal L$. The ground states are \emph{locally indistinguishable} (see, e.g., Ref.~\cite{my_proof}), by which we mean that the reduced density matrix, $\rho_{A} = {\rm Tr}_{\bar A} \ket{\Psi_{ij}} \bra{\Psi_{ij}}$, is the same for all ground states as long as the distinguished region $A$ does not form a spanning cluster in the lattice and it is smaller than its complement ${\bar A}$.


\section{Wavefunction Deformation \label{sec:III}}
We add a fine-tuned random perturbation to the toric code, which is controlled by a set of parameters $\{\beta_j\}$, $j=1,\ldots,N$. In this case the Hamiltonian is given by
\begin{eqnarray} \label{RFCC}
H = H_{\rm TC} + \sum_{s} \exp \left(- \sum_{j\in s} \beta_j \hat{\sigma}^{\rm z}_{j} \right),
\end{eqnarray}
where the first sum runs over all stars and the second sum runs over the four spins in each star. For small values of $\beta_{j}$ the perturbation term is equivalent to a local magnetic field. The ground state can be obtained as in Refs.~\cite{Castelnovo_Chamon,Stefanos}, where the clean-case equivalent was studied ($\beta_j = \beta, \; \forall \; j\in {\cal L}$). It was shown there that the model has a quantum critical point at $\beta_c \simeq 0.4406868$, and topological order is lost for $\beta > \beta_c$.

Without detailing the derivation \cite{Castelnovo_Chamon} (as it is not relevant for our purposes here), the ground state of the disordered model~\eqref{RFCC} is explicitly given by
\bea
\ket{\Psi_{0}} = {\cal Z}^{-1/2} \sum_{g\in G} \exp\left[\frac{1}{2}\sum_{j} \beta_j \sigma^{\rm z}_j(g) \right] g \ket{\bf 0},
\eea
where
\bea
{\cal Z} \equiv \sum_{g\in G} \exp \left[\sum_{j} \beta_j \sigma^{\rm z}_j(g) \right].
\eea

\subsection{Mapping to a Classical Random-Bond Ising Model}
A generic configuration $g \ket{\bf 0}$ is uniquely determined by the set of star operators acting on the reference state $\ket{\bf 0}$, modulo the action of the product of all $g\in G$, which is equal to the identity. Therefore, there is a $1$-to-$2$ mapping between $G$ and the configuration space $\Theta = \{ \theta \}$ of an \emph{Ising model} with degrees of freedom $\theta^{\ }_{s}$ living on the sites $s$ of the square lattice. We change variables from the bonds ($\sigma$ spins) of the lattice to the vertices (stars or $\theta$ spins) by choosing $\theta_{s} = -1$ ($+1$) to mean that the corresponding star operator $A_s$ is (is not) `participating' in the relevant $g\in G$. Since each $\sigma$-spin can only be flipped by its two neighboring $\theta$-spins, we have $\sigma_{j} \equiv \theta_{s}\theta_{s^{\prime}}$, where $j$ labels the bond between two neighboring sites $\langle s,s^{\prime} \rangle$. Hence, each random parameter $\beta_j$ is uniquely labeled by the pair $s,s^{\prime}$, and we can substitute $\beta_j \rightarrow \beta_{ss^{\prime}}$.

With this change of variables, the ground state of $H$ can be expressed as
\begin{eqnarray} \label{Psi_0_theta}
\ket{\Psi_0} = \sum_{\theta \in \Theta}
    \frac{1}{\sqrt{\cal Z}}
    \exp \left( \frac{1}{2} \sum_{\langle s,s^{\prime} \rangle}
          \beta_{ss^{\prime}} \theta_{s} \theta_{s^{\prime}} \right)
      g(\theta) \ket{\bf 0},
\end{eqnarray}
where
\begin{equation} \label{Z_theta}
{\cal Z} = \sum_{\theta \in \Theta}
  \exp \left( \sum_{\langle s,s^{\prime} \rangle}
     \beta_{ss^{\prime}} \theta_{s} \theta_{s^{\prime}} \right).
\end{equation}
The second quantity is, in fact, the partition function of a classical random-bond Ising model (RBIM) on a 2D square lattice at temperature $T$, with reduced nearest-neighbor couplings $\beta_{ss^{\prime}} = J_{ss^{\prime}}/T$. As a result, all equal-time correlation functions that can be expressed in terms of $\theta_{s}$ variables are exactly the same in the two systems.

\begin{figure}[ht]
\includegraphics[width=0.95\columnwidth]{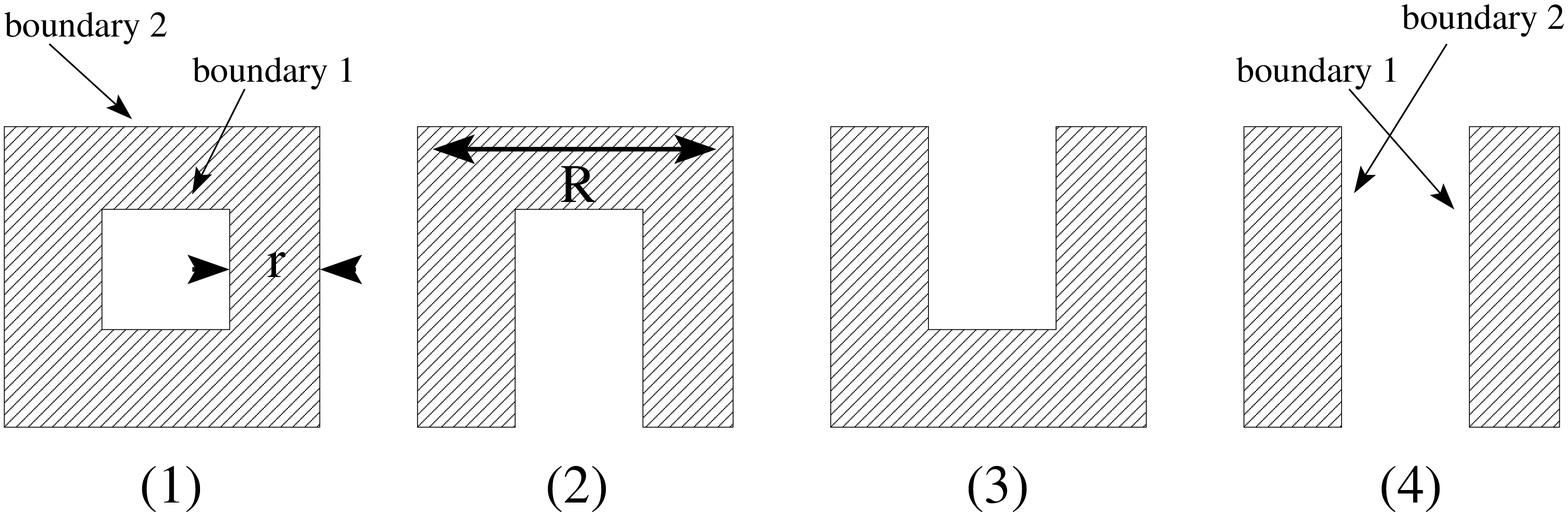}
\caption{
\label{fig: topological partitions}
The four bipartitions $({\A}_{i},{\B}_{i})$ used to compute the topological entropy according to Levin and Wen \cite{S_topo}, $S_{\rm topo} = \lim_{r,R\to\infty} [- S({\A}_{1}) + S({\A}_{2}) + S({\A}_{3}) - S({\A}_{4})]$,
where $S(\A) = - {\rm Tr}(\rho_{\A} \ln \rho_{\A})$ is the von Neumann entropy of subsystem $\A$.}
\end{figure}
%

\subsection{Topological Entropy and Phase Diagram}
We can now calculate the entanglement and topological entropy~\cite{S_topo}, which serves as an order parameter for phase transitions into and out of topological order~\cite{Castelnovo_Chamon, my_proof, Castelnovo2007, Hamma_detectionTQPT}. Via the topological entropy the phase diagram of model (\ref{RFCC}) can be obtained exactly by identifying the correspondence between topological order and spin polarized phases with the paramagnetic and ferromagnetic phases of the mapped classical model, as explained now.

Using the bipartition scheme of Levin and Wen \cite{S_topo}, depicted in Fig.~\ref{fig: topological partitions}, a calculation following closely the steps of Ref.~\cite{Castelnovo_Chamon} leads to the topological entropy
\bea
S_{\textrm{topo}} = \lim_{r, R \to \infty} \frac{1}{\cal Z}
 \sum_{g \in G} e^{\sum_{\langle s,s' \rangle} \beta_{ss'} \theta_{s} \theta_{s^{\prime}} } \log_{2} R(g)
\eea
where
\begin{equation}
R(g) \equiv \frac{\left[ {\cal Z}^{\partial}_{1}(g) + {\cal Z}^{\partial,\,\textrm{twisted}}_{1}(g) \right] \left[ {\cal Z}^{\partial}_{4}(g) + {\cal Z}^{\partial,\,\textrm{twisted}}_{4}(g) \right]} {{\cal Z}^{\partial}_{2}(g) Z^{\partial}_{3}(g)}.
\label{eq: formula for the topo entropy}
\end{equation}
Here ${\cal Z}^{\partial}_{2,3}(g)$ represents the partition function of an Ising model with reduced nearest-neighbor interactions $\{\beta_{ss^\prime}\}$, and with fixed spins along the boundary of bipartitions $2$ and $3$, respectively (see Fig.~\ref{fig: topological partitions}). Partition functions ${\cal Z}^{\partial}_{1,4}(g)$ are the analogs of ${\cal Z}^{\partial}_{2,3}(g)$ for bipartitions $1$ and $4$, respectively, whereas ${\cal Z}^{\partial,\,\textrm{twisted}}_{1,4}(g)$ differ from them in that the spins along one of the two boundaries are flipped.

The sum over $g$ in $S_{\rm topo}$ acts as a weighed average of the logarithmic term over all possible values of the spins at the boundary. Note that in Eq.~\eqref{eq: formula for the topo entropy} the partitions with two boundaries, and hence with non-trivial topology, are those appearing with two contributions (i.e., bipartitions 1 and 4). These contributions are responsible for a non-vanishing $S_{\rm topo}$.

In the high-$T$ phase of the RBIM ($\beta_{ss^\prime} \to 0$) the correlations are short ranged and the choice of boundary conditions affects the partition function with only exponentially small corrections. Therefore, in this case,
\bea
{\cal Z}^{\partial}_{1}(g)Z^{\partial}_{4}(g)
\simeq
{\cal Z}^{\partial,\,\textrm{twisted}}_{1}(g)Z^{\partial}_{4}(g)
\simeq
\,\ldots\,
\simeq
{\cal Z}^{\partial}_{2}(g) Z^{\partial}_{3}(g) \nonumber
\eea
and $S_{\textrm{topo}} = 2$. This corresponds to the topological phase of the original model~\eqref{RFCC}. On the contrary, in a ferromagnetically ordered phase, e.g., when $\beta_{ss^\prime}=\beta \gg 1$, we have
\bea
{\cal Z}^{\partial}_{1}(g)
\gg
{\cal Z}^{\partial,\,\textrm{twisted}}_{1}(g),
{\cal Z}^{\partial}_{4}(g)
\gg
{\cal Z}^{\partial,\,\textrm{twisted}}_{4}(g), \nonumber
\eea
while
$
{\cal Z}^{\partial}_{1}(g)Z^{\partial}_{4}(g)
\simeq
{\cal Z}^{\partial}_{2}(g) Z^{\partial}_{3}(g)
$
still holds~\cite{Castelnovo_Chamon}. This leads to $S_{\textrm{topo}} = 0$. The behavior of $S_{\rm topo}$ across such a transition is discontinuous, with a sharp jump from $S_{\textrm{topo}} = 2$ to $S_{\textrm{topo}} = 0$.

\begin{figure}[ht]
\includegraphics[width=0.85\columnwidth]{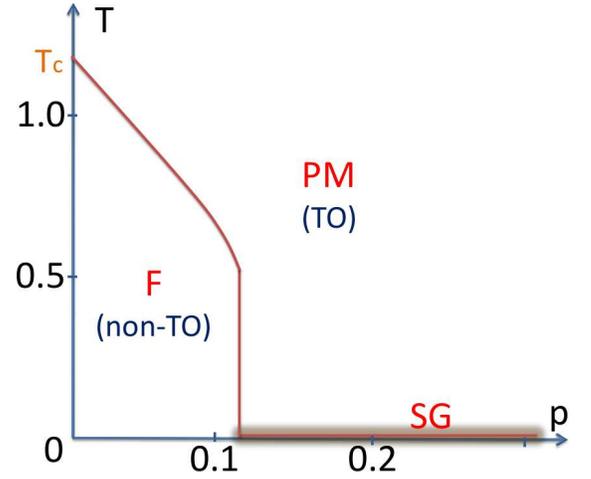}
\caption{
\label{fig:phase diagram1}
(Color online) Phase diagram the random-bond Ising model \cite{RBIM} (RBIM) with bipartite disorder as a function of the density of antiferromagnetic (AF) bonds ${\rm p}$ and $T \equiv 1/\beta$.
}
\end{figure}
%

\subsection{Bipartite Disorder}
Let us study a particular case of disorder in model~\eqref{RFCC}, with probability distribution
\bea
\mathcal{P}(\beta_{ss^{\prime}}) = {\rm p} \delta\left(\beta_{ss^{\prime}} + \frac{1}{T} \right) + \left(1-{\rm p} \right) \delta\left(\beta_{ss^{\prime}} - \frac{1}{T} \right).
\eea
The ground state of the model, Eq.~\eqref{Psi_0_theta}, is controlled by the partition function ${\cal Z}$ of a 2D RBIM with reduced couplings $\beta_{ss^{\prime}} = \pm 1/T$, where the density of antiferromagnetic (AF) bonds is ${\rm p}$. The phase diagram of this model is well-known~\cite{RBIM,RBIM_3d} and is given for completeness in Fig.~\ref{fig:phase diagram1}. The solid line separates the paramagnetic (PM) and ferromagnetic (F) phases. In 2D a spin glass (SG) phase appears only in the zero temperature limit (i.e., for $\beta_{ss'} \to \infty$). It is noted that other salient features in the phase diagram are straightaway inherited from the physics of spin glasses. For instance, model~\eqref{RFCC} exhibits a multicritical Nishimori point and a re-entrant behavior in the FM phase in Fig.~\ref{fig:phase diagram1} (see, e.g., Ref.~\cite{RBIM}, and references therein).

The region of the phase diagram where the RBIM is in the paramagnetic phase corresponds to a topologically ordered phase of the quantum model, which is continuously connected to the unperturbed toric code. In this fine tuned model, topological order is stabilized by disorder, and for instance at ${\rm p}=1/2$ topological order survives \emph{for all} non-vanishing values of $T$. The mapped system enters the spin glass phase only for $T \to 0$. However, specifically in 2D, the stiffness exponent of the spin glass phase is negative~\cite{Bray1984} and this leads to $R = 1 \; \rightarrow \; S_{\textrm{topo}} = 2$ down to $T=0$, while the system becomes conformally invariant~\cite{Hastings2006}.

It is remarkable that if we prepare the system in the spin polarized phase, which can be done in a straightforward way by applying a strong uniform field $\beta_{j} = \beta > \beta_c$, and then we increase the amount of disorder (that is, if we increase the density of AF bonds ${\rm p}$), we can \emph{induce a recovery of topological order} in the disordered model of Eq.~\eqref{RFCC}.


\section{Hamiltonian Deformation \label{sec:IV}}
We now focus on the more general case of the toric code in a random magnetic field,
\begin{equation} \label{Random_TC}
{\cal H} = H_{\rm TC} - \sum_{i \in {\cal L}} h_i \hat{\sigma}^{z}_{i}.
\end{equation}
In the limit $\vert h_i \vert \to \infty$ every spin $i$ polarizes in the local field direction and hence the ground state has no topological order. Contrary to the uniform-field case, such a strong random field can favor plaquettes for which $B_p = -1$. An upper bound to the stability of this fully polarized phase can be obtained by setting $\lambda_A = 0$ and asking how large $\lambda_B$ needs to be to make it energetically favorable for some spins to point against the field direction so that $B_p = \prod_{i \in p} \sigma^z_i = +1$ everywhere on the lattice. Recall that negative plaquettes can only be removed in pairs by flipping strings of spins that join them on the dual lattice. Therefore, the energy gain for such a change, $-4\lambda_B$, is opposed by a field contribution $2\sum \vert h_i \vert$, where the sum is carried out along the chosen string. If the probability of $h_i$ being positive is the same as that of being negative, the density of negative plaquettes is $\eta = 1/2$ and the typical separation between two such defects is $\ell \sim 1/\sqrt{\eta}$, independently of system size. The field contribution amounts to $2 \ell \overline{h} \sim 2 \overline{h} / \sqrt{\eta}$, where $\overline{h}$ is the average field strength (extremal field values need not be considered here as it is almost always possible to find a string that avoids them on the lattice). Once $\lambda_B \gtrsim \overline{h} / 2\eta$, the condition $B_p=+1$ becomes enforced throughout the lattice and the system exits the fully polarized phase.

\subsection{Mapping to a Quantum Random-Bond Ising Model}
So long as $\lambda_B \gtrsim \overline{h} / 2\eta$, the phase diagram can be studied by considering the low-energy effective Hilbert space given by the span of all $\sigma^z$ tensor product states satisfying the condition $B_p = +1$, as in Ref.~\cite{Trebst}. We can then simplify Hamiltonian~\eqref{Random_TC} by introducing operators $\hat{\theta}^{z}_s$ living on the sites of the lattice, with eigenvalues $\theta^{z}_s \equiv (-1)^{n_s} = \pm 1$, where $n_s$ is the number of times that the four spins around $s$ have been flipped with respect to some reference configuration, say the fully magnetized state ($\sigma^z_i = +1$). The value of $\sigma^z_i$ for a spin at bond $i$ is uniquely determined by the product of two $\theta^z_s$ on adjacent sites, $\sigma^z_i = \theta^z_s \theta^z_{s^\prime}$. Indeed, in the restricted Hilbert space, a spin can only be flipped as part of a $4$-spin flip process around one of the two adjacent stars (otherwise it leads to a high energy state $\sim 2 \lambda_B$.) If we interpret $\hat{\theta}^z_s$ as the $z$-component of a spin-$1/2$ degree of freedom, the operator $A_s$ is the corresponding $x$-component. Therefore, in terms of the $\theta$-spins, the Hamiltonian ${\cal H}$ can be rewritten as
\bea
{\cal H} = - \lambda_A \sum_s \hat{\theta}^x_s - \sum_{\langle s,s^{\prime} \rangle} h_{ss^\prime} \hat{\theta}^z_{s} \hat{\theta}^z_{s^\prime} ,
\label{eq: TRBIM}
\eea
where in the second term we rename $i \leftrightarrow ss'$. This is the transverse field RBIM on the square lattice, whose phase diagram has been studied for several instances of disorder, such as $\pm r$ (bipartite) bond disorder \cite{dosSantos1984} and bond-diluted disorder \cite{Stinchcombe1981}. The phase diagram of the first case is shown in Fig.~\ref{fig:phase diagram2}, while that of the second is shown in Fig.~\ref{fig:SM}. Each case is studied in turn in the following subsections.

\subsection{\label{sec: bond dilution}
Bond Dilution Disorder and the Percolation Threshold
           }
Let us start for simplicity by considering the Hamiltonian~\eqref{Random_TC} for a toric code in a random local magnetic field, with bond-diluted disorder according to the probability distribution
\begin{equation}
\mathcal{P}(h_i) = (1-{\rm p}) \delta(h_i) + {\rm p} \delta(h_i - h).
\end{equation}
In other words, a uniform magnetic field of strength $h$ is applied to a fraction $\rm p$ of $\sigma$ spins chosen at random.

Contrary to the $\pm r$ case, the dual mapping to $\theta$ spins holds for all values of $h$, since the diluted field does \emph{not} compete with the plaquette term $B_p$. In the dual language of Eq.~\eqref{eq: TRBIM} this type of disorder leads to a bond-diluted transverse RBIM.

\begin{figure}[ht]
\includegraphics[width=0.85\columnwidth]{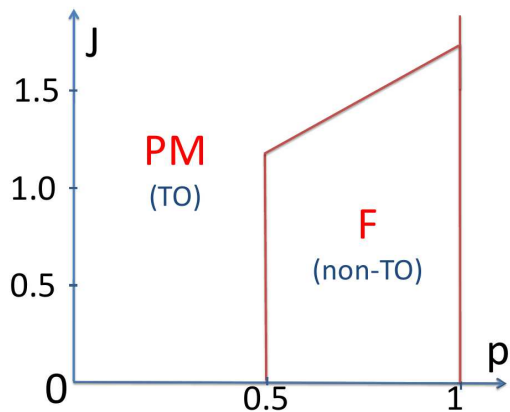}
\caption{
\label{fig:SM}
(Color online) Phase diagram of the random-bond Ising model (RBIM) with bond-diluted disorder~\cite{Stinchcombe1981} ($J\equiv \lambda_A / h$).
}
\end{figure}

The $h$-${\rm p}$ phase diagram of this model is known~\cite{Stinchcombe1981} and shown here in Fig. \ref{fig:SM}. For bond dilution lower than a critical value ($\rm p \ge \rm p_{\rm c} = 1/2$), the system exhibits a quantum phase transition from a paramagnetic (PM) phase for small values of $h/\lambda_A$ to a ferromagnetic phase for large values of $h/\lambda_A$. For bond dilution above the critical value, the phase transition is suppressed altogether and the system remains in the PM phase for any $h$.

In the language of the original model~\eqref{Random_TC}, the PM phase maps onto a topologically ordered ground state. Therefore topological order survives irrespective of field strength all the way up to the critical dilution, i.e., for $0 \le \rm p \le \rm p_{\rm c}$. Beyond this threshold, a sufficiently large field takes the system into a spin polarized phase.

This result can be understood using a percolation argument applied to the large $h$ case where each spin $\sigma_i$ with $h_i \neq 0$ is polarized in the direction of the field. The Hilbert space of the system is thus effectively reduced to that of the remaining $(1-{\rm p})N$ spins. Below the percolation threshold ($\rm p_{\rm c} = 1/2$) all clusters of polarized spins are finite. One can verify that the corresponding projection in the large $h$ limit simply changes the lattice on which the toric code is defined, but does not alter its topological properties. From a quantum information perspective, this implies a high ($50\%$) error threshold~\cite{Dennis,Sean_Barrett}.

On the contrary, for ${\rm p} \geq \rm p_{\rm c}$ a spanning cluster appears. As $h$ is increased and the spins on the cluster polarize, the projected system breaks down into disconnected components and topological order is lost. For instance, winding loops are bound to intersect the percolating cluster and topological sectors can no longer be degenerate (as discussed in the case of bipartite disorder with $\pm r$ magnetic field, in the previous subsection).

%

\subsection{Bipartite Disorder and the Topological Spin Glass}
In the $\pm r$ model~\cite{dosSantos1984}, a uniform field of strength $r$ is applied to all $\sigma$-spins but its direction is randomly inverted for a fraction $\rm p$ of the spins. Mathematically, the relevant probability distribution is
\bea
\mathcal{P}(h_i) = \left(1-{\rm p}\right) \delta\left(h_i + r \right) + {\rm p} \delta\left(h_i - r \right).
\eea

Topological order corresponds to the fixed point $J \equiv \lambda_{\rm A}/r \to \infty$ (or, equivalently, to the $\theta^x$-polarized PM phase). It is stable for all values of AF-bond density ${\rm p}$, provided the disordered field strength obeys $r \ll \lambda_A$.

The case of uniform field (i.e., the line ${\rm p}=1$) yields the 2D transverse field Ising model, whose phase diagram is controlled by the 3D classical Ising model~\cite{book_chapter}. The ferromagnetically ordered phase, which corresponds to a non-topological phase in model~\eqref{Random_TC}, is controlled by the $J=0,\; {\rm p}=1$ fixed point, and it is stable in a finite portion of the phase diagram.

\begin{figure}[ht]
\includegraphics[width=0.90\columnwidth]{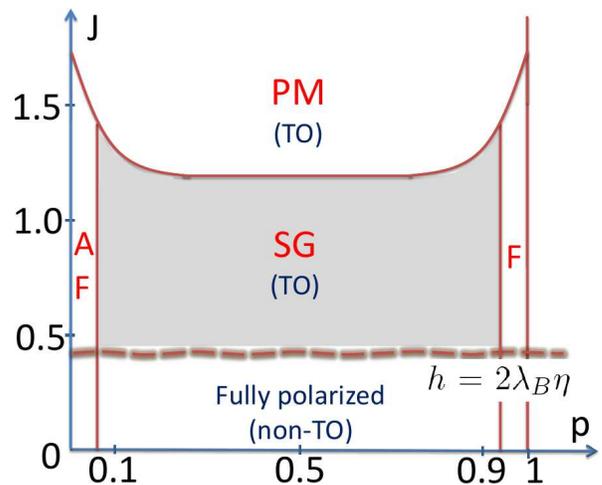}
\caption{
\label{fig:phase diagram2}
(Color online) Phase diagram of the transverse field RBIM \cite{dosSantos1984} as a function of ${\rm p}$ and $J \equiv \lambda_A / r$. A \emph{topological spin glass} [in the original model~\eqref{Random_TC}] appears, indicated by the shaded region.
}
\end{figure}

There is a third region in the phase diagram in Fig.~\ref{fig:phase diagram2}, controlled by the $J=0, \; {\rm p}=1/2$ fixed point. It occurs when the field strength $r$ is large with respect to the transverse field $\lambda_{\rm A}$ (i.e., $J$ is small) and yet a high density of AF bonds ${\rm p}$ prevents the onset of ferromagnetic order. In this case the system enters a quantum spin glass phase~\cite{dosSantos1984}.

Contrary to model~\eqref{RFCC}, the fate of topological order across this glass transition is difficult to determine via a calculation of the topological entropy of the system. Here we take an alternative route and propose that this is in fact a new phase of matter where topological order \emph{coexists} with spin glass properties.

We consider for simplicity the case ${\rm p} = 1/2$, where Eq.~\eqref{eq: TRBIM} reduces to the transverse field bimodal Edwards-Anderson model. In the weak transverse field limit $\lambda_{\rm A} / r \ll 1$ it is customary to construct approximate ground state wavefunctions by taking the tensor product states that minimize the classical EAB energy (Eq.~\eqref{eq: TRBIM} with $\lambda_{\rm A} = 0$) and forming superpositions of any sets of such states that are connected by the action of $\hat{\theta}^x_s$ operators, $s \in {\cal L}$. These ground states were shown to involve an extensive number of tensor product states and they exhibit statistical properties similar to the PM state~\cite{Landry2004} (i.e., the topologically ordered state in the dual $\sigma$ variables).

The ensemble of all the bonds in the EAB model that are always satisfied or always frustrated in all ground state configurations is called the \emph{rigid lattice}~\cite{Barahona1982}. Clusters of $\theta$ spins connected by bonds in the rigid lattice are locked with respect to one another. Correspondingly, the original $\sigma$ spins associated with the bonds of the rigid lattice are pinned to a fixed value in the ground state configurations.

As we showed in Sec.~\ref{sec: bond dilution} in the case of a diluted field, pinning the value of spins at random in the toric code model does not spoil its topological properties until we pin a percolating cluster of them. As a result, we see that the survival of topological order in the spin glass phase of the toric code model in a bimodal field is related to the percolation properties of the rigid lattice in the Edwards-Anderson bimodal spin glass model.

Very recent numerical results on the EAB model~\cite{Roma2010} suggest that the rigid lattice does \emph{not} percolate in 2D, although the results cannot be deemed conclusive. If they are confirmed~\cite{footnote2}, our results suggest that topological order in model~\eqref{Random_TC} survives inside the quantum spin glass phase. This occurs for field strengths $r$ stronger than $\lambda_A$ but below the threshold value $2\lambda_B\eta$, and high AF bond densities ${\rm p}$. As $r$ is further increased beyond $2\lambda_B\eta$, the system eventually exits the topological spin glass phase and enters the field polarized phase discussed earlier.

\section{Comparison between Wavefunction and Hamiltonian Deformations \label{sec:V}}
One can develop a qualitative understanding of the difference between models~\eqref{RFCC} and~\eqref{Random_TC} by studying more closely the two terms they differ by, that is, $\sum_s\exp\left[-\sum_{i \in s} h_i \sigma^z_i\right]$ and $-\sum_{i} h_i \sigma^z_i$, respectively. We shall consider the case $h_i = \pm h$ for convenience.

The term in model~\eqref{Random_TC} is a local magnetic field which favors a unique tensor product state with an energy gap $\geq 2h$ to any other state. When added to the toric code, which has a finite gap, this term eventually dominates at large field strength and the ground state tends to the field-polarized state, where topological order is lost.

While the perturbation of model~\eqref{RFCC} favors precisely the same state, it does so in a subtler way, as it becomes evident when we consider its action on a single star $s$ composed of four spin-$1/2$ degrees of freedom, $\exp\left[-\sum_{i \in s} h_i \sigma^z_i\right]$. In this case, the lowest energy state where each spin points in the local field direction has a gap to the first excited state proportional to $e^{-2h} - e^{-4h}$.

In contrast to model~\eqref{Random_TC}, the energy difference now tends to zero as the field strength $h$ is increased. (Note that this is not true for higher excited states, whose energy difference with respect to the lowest state, $e^{2h} - e^{-4h}$ and $e^{4h} - e^{-4h}$, diverges exponentially with $h$).
As a result, there are exponentially many tensor product states within an energy band smaller than the gap of the unperturbed toric code irrespectively of the value of $h$.

This argument illustrates why topological order cannot survive for
large values of $h$ in~\eqref{Random_TC}, while the perturbation
in~\eqref{RFCC} is by far less disruptive. Nevertheless, the
survival of topological order for all values of $h$
in~\eqref{RFCC} is a non-trivial result that cannot be
straightforwardly deduced from these simple considerations.


\section{Discussion and Perspectives \label{sec:VI}}
We have shown that topologically ordered quantum systems respond to quenched disorder very differently from conventionally ordered systems~\cite{footnote}, in the following sense. In general, close to a transition out of a topologically ordered phase, which is inherently long-range entangled, the introduction of local disorder in the applied field stabilizes the topological phase. In some cases, such as for random field dilution of sufficiently high density, topological order is robust with respect to any strength of the external magnetic field.

An intuitive framework in which the main results can be understood on a physical level may be provided by the quasiparticle excitations (anyons) in the system. A uniform field can be seen as a hopping term for otherwise static anyons, allowing them to propagate with some lattice momenta. For small fields the system has a spectral gap and retains topological order as the ground state is adiabatically connected with that of the toric code~\cite{Hastings}. For large fields the gap closes and topological order is lost to anyons winding around the torus. It is then tempting to speculate that a disordered field takes on the role of either a pinning potential or a random hopping term, which \emph{localizes} the anyons. Such a mechanism would indeed preserve topological order, as it denies the propagation of pair-created anyons around the torus.

We also argued that quenched disorder in topologically ordered systems might lead to a new quantum phase where topological order and spin glassiness coexist. Fully understanding the complex nature of this intermediate quantum phase requires the combined expertise of both research areas. In particular, our results lead naturally to the following set of questions, which deserves further study.

Spin glass transitions are usually detected via magnetic susceptibility measurements. In the present context, these involve an odd number of $\theta$-spins, i.e., they are intrinsically non-local in the original $\sigma$-spin degrees of freedom. Are there any local observables in the original model that can detect the topological spin glass transition? What properties characterize the new phase? Do long relaxation timescales appear in the quasiparticle dynamics of a topological spin glass~\cite{Chamon2005}? Could the glassy nature of this intermediate quantum phase be exploited to engineer a topological quantum memory that is robust against thermal fluctuations~\cite{Hamma2009}?

In conclusion, new phenomena arise at the confluence of topological order and glassy physics, with potentially interesting implications for both areas and various useful applications in quantum information science.


\section*{Acknowledgments}
CC is deeply indebted to C.~Chamon and C.~Mudry for insightful discussions. We thank R.~Stinchcombe and S.~L.~A.~de~Queiroz for discussions on classical and quantum RBIMs, and J. Eisert for useful comments. Financial support by EPSRC-GB Grant Nos. EP/G045771/1 (DIT, TJO) and EP/G049394/1 (CC) is acknowledged.



\begin{thebibliography}{99}
\bibitem{footnote}
    While we focused on the case of $\pm r$ fields and field dilution, similar results are expected to hold for other distributions, such as Gaussian and box disorder. Moreover, it is known that the properties of the perturbed toric code depend on the direction of the magnetic field~\cite{Vidal}. It will be interesting to see how the behavior changes in a random field along the $y$ direction.
\end{thebibliography}
\end{document}